\documentclass{article}  
\usepackage{spadre2008}
\usepackage{graphicx}
\usepackage{wrapfig}

\def\rightmark{Superhorizon Fluctuations}
\def\leftmark{P. Sorensen}
 
\title{Searching for Superhorizon Fluctuations in Heavy-Ion Collisions } 
\authors{
{Paul Sorensen %
}\\[2.812mm]
{\normalsize
\hspace*{-8pt} Brookhaven National Laboratory, \\ 
Upton NY, USA\\[0.2ex] 
}}
 
\abstract{ In this talk I discuss novel explanations for the azimuthal
  correlations observed in heavy-ion collisions. I review some ideas
  about correlations and the evolution of heavy-ion collisions. Some
  aspects of the correlations observed in heavy-ion collisions may be
  indicitive of the suppression of super-horizon fluctuations. }

\keyword{super-horizon, fluctuations, correlations, elliptic flow } 
\PACS{25.75.-q}
  
\begin{document}

\maketitle
\setcounter{page}{1}

\section{Introduction}\label{intro}

It has been argued that the data from RHIC experiments indicate that
the matter created in heavy-ion collisions behaves as a nearly perfect
liquid with viscosity near a lower bound predicted by string theory
and by quantum mechanics. These conclusions are based largely on comparisons
of hydrodynamic models to the shape of single particle spectra
and $v_2$ for particles of different masses~\cite{wp}. These conclusions depend
crucially on the initial conditions assumed for the models,
particularly the initial eccentricity~\cite{ic}. Here, I show that the upper
limit on $v_2$ fluctuations coincides with the fluctuations expected
from Monte Carlo Glauber models of the initial eccentricity
fluctuations.

The upper limit on $v_2$ fluctuations is derived without subtracting
contributions from two-particle correlations unrelated to the
reaction-plane. As such, either
\begin{quote}
1) Those correlations invalidate the models for the initial
eccentricity,

2) the observed two-particle correlations are a manifestation of the
correlations and fluctuations in the initial source,

3) or both 1) and 2) are true.
\end{quote}
The structure of the correlations expected from the initial conditions
will depend on the details of the expansion of the fireball: e.g. the
formation time, the lifetime, the characteristic size of the fireball,
and the characteristic size of the correlations. In
Ref.~\cite{Mishra:2007tw}, Mishra et. al. argue that the longest wavelength
fluctuations from the initial conditions of heavy-ion collisions may
remain super-horizon throughout the evolution of the fireball. In this
case, the long wavelength modes should be suppressed. I show that the
suppression of these modes can explain a negative contribution to
$p_T$ correlations which was previously attributed to a recoil caused
by a fast parton impinging on the medium. I argue that the strong
correlations seen in heavy-ion collisions may not actually reflect
correlations from fragmenting quarks or gluons (mini-jets) but rather
may reflect correlations from lumpy initial conditions or from cluster
formation at a phase boundary.

\section{$v_2$ Fluctuations and Initial Conditions}

Elliptic flow ($v_2$) measurements are sensitive to the shape of the
initial overlap zone so $v_2$ fluctuations can reveal
information about fluctuations and correlations in the initial
geometry. Distinguishing between $v_2$ fluctuations $\sigma_{v_2}^2$
and non-reaction-plane correlations $\delta_2$ (non-flow) requires
knowledge of the reaction-plane or information about the
$\Delta\phi$-, $\Delta\eta$-, charge-sign- or multiplicity-dependence
of non-flow. Lacking this information, only the sum of non-flow and
$v_2$ fluctuations can be determined $\delta+2\sigma_{v_2}^{2}$. Then
the upper limit on $\sigma_{v_2}/\langle v_2 \rangle$ can be
calculated and compared to models of eccentricity fluctuations~\cite{me,phobos}.

\begin{figure}[htb]
\centering\mbox{
\vspace{-10pt}
\includegraphics[width=0.5\textwidth]{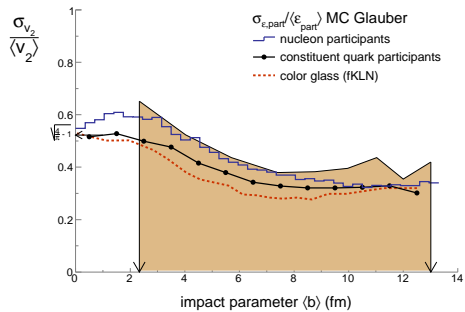}}
\vspace{-0pt}
\caption[]{  Upper limit on $\sigma_{v_{2}}/\langle
   v_{2}\rangle$ compared to models of eccentricity.}
\label{f2}
\end{figure}

Fig.~\ref{f2} shows the upper limit on $\sigma_{v_2}/ \langle
v_2\rangle$ compared to several Monte Carlo models of
$\varepsilon_{part}$ fluctuations. All models lie within the allowed
range while the CGC model~\cite{cgc} and the model based on
constituent quarks inside the participating nuclei both have smaller
relative widths. The nucleon participant model leaves little room for
other sources of fluctuations and correlations beyond the initial
geometric ones. The large near-side peak observed in two-particle
correlations~\cite{ridge} contradict the idea that all or most of
$\delta+2\sigma_{v_2}^2$ is dominated by $\sigma_{v_2}$ suggesting
that the CGC or constituent-quark model may be preferred. Recently it
was proposed that correlations and fluctuations in the initial
conditions may also contribute to the near-side
ridge~\cite{Dumitru}. Either the density fluctuations in the initial
state are manifested in both the near-side peak and in
$\sigma_{v_{2}}$, or the initial overlap density is smoother than
would be expected from the MC Glauber model. The translation of
fluctuations from the initial conditions into observed correlations
has analogies with the expansion of the universe and the temperature
fluctuations in the CMB.

\section{Heavy Ion Expansion: It's About Time}

\begin{figure}[htb]
\centering\mbox{
\vspace{-15pt}
\includegraphics[width=0.8\textwidth]{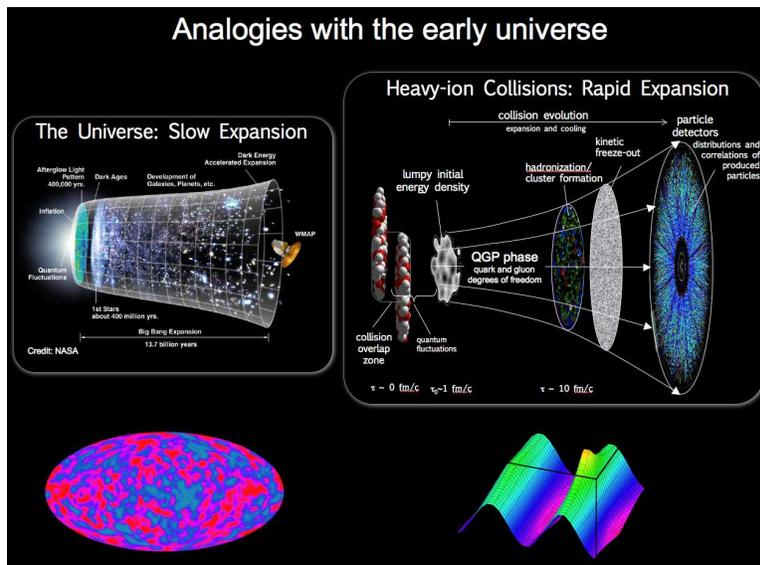}}
\vspace{-15pt}
\caption[]{ This figure presents an analogy between the expansion of
  the universe and the expansion of the fireball created in heavy-ion
  collisions. The graphic on the bottom right shows $p_T$ correlations
  measured by STAR which are analogous to the Temperature fluctuations
  shown on the bottom left measured by WMAP~\cite{Komatsu:2008hk}. }
\label{bigfig}
\end{figure}

Measurements of the scale dependence of temperature fluctuations
support the cosmic inflation scenario of cosmology. In this scenario,
the temperature fluctuations in the cosmic microwave background
radiation result from quantum fluctuations that are magnified during
an inflationary epoch by a factor of $\approx 10^{30}$. The fireball
created in heavy-ion collisions may last for as short as 10 fm. In
this case, long-wavelength correlations in the transverse direction
can remain super-horizon throughout the fireballs evolution and would
therefore be suppressed in the observed power spectrum. Causality in
the longitudinal plane has been discussed previously in~\cite{Dumitru}
and references therein.

\subsection{Yea, Though I Walk Through the Valley}

\begin{figure}[htb]
\vspace{-5pt}
\hspace{0.1\textwidth}
\resizebox{0.33\textwidth}{!}{\includegraphics{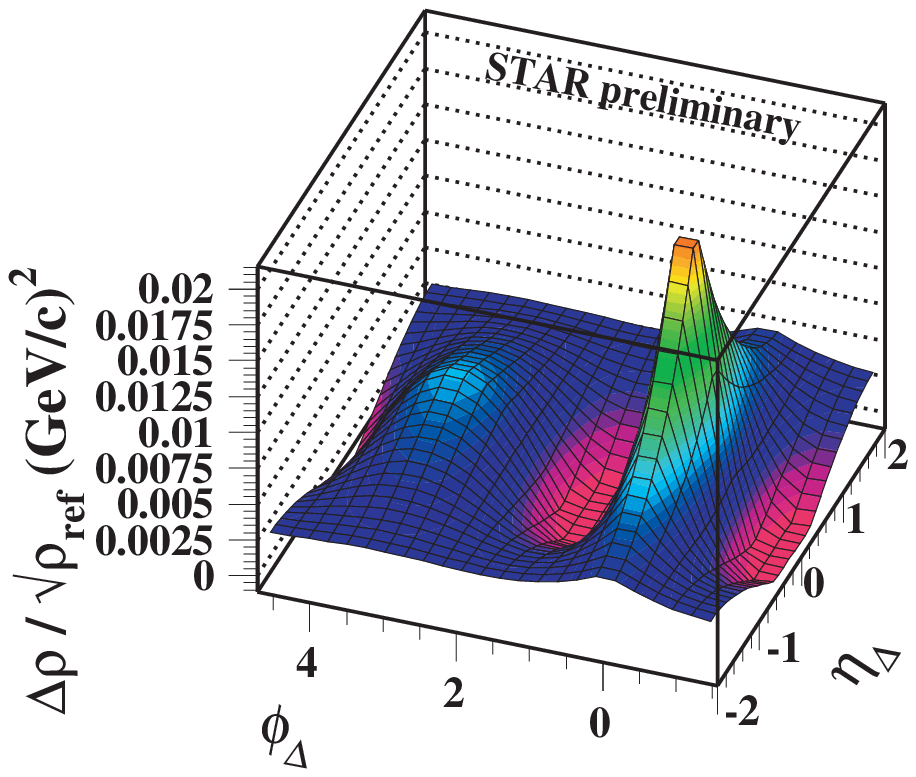}}
\hspace{0.15\textwidth}
\resizebox{0.22\textwidth}{!}{\includegraphics[width=0.5in,height=0.4in,angle=90]{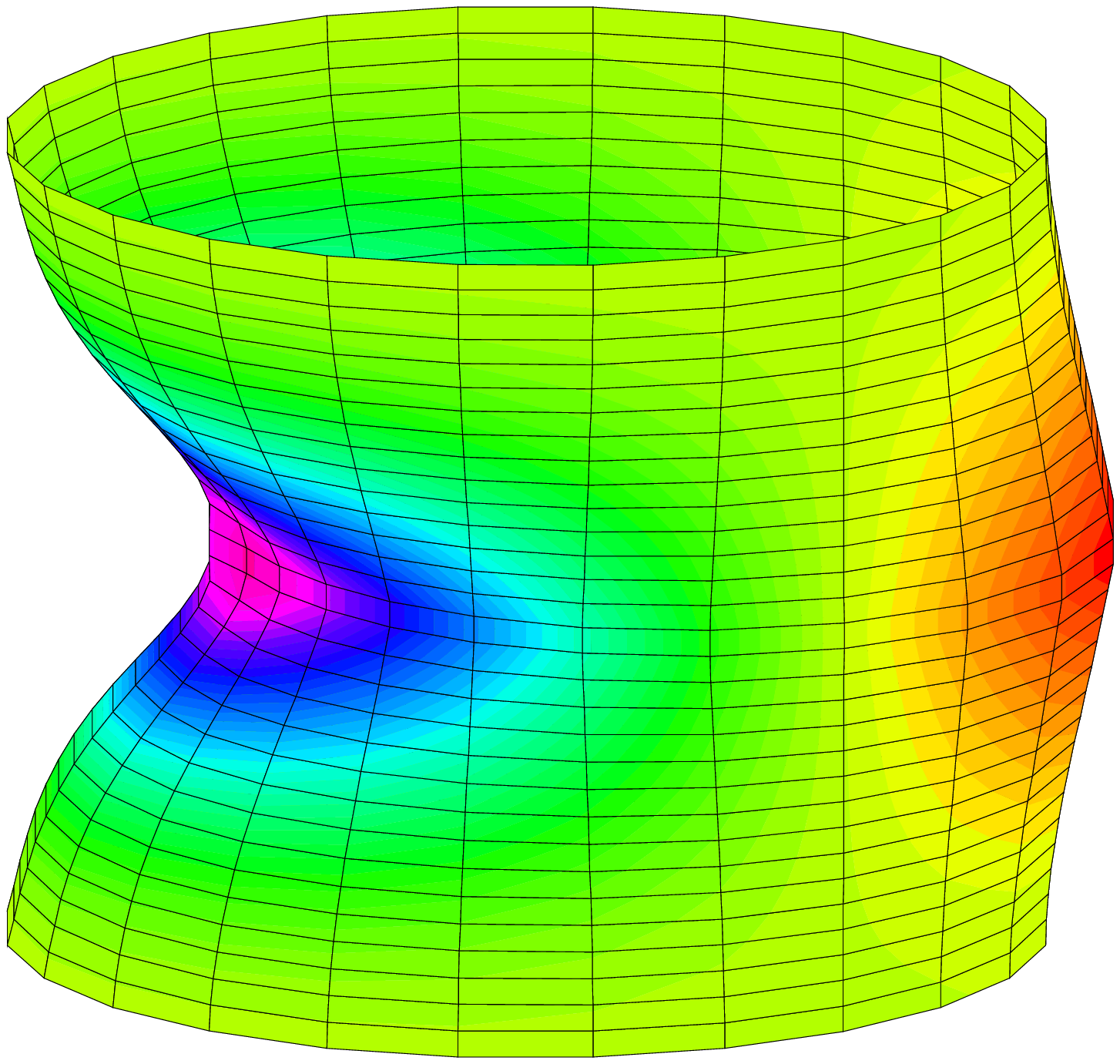}}
\vspace{-15pt}
\caption[]{ Left panel: $p_T$ autocorrelations derived from the
  $\langle p_T \rangle$ fluctuation scale dependence in $Au+Au$
  collisions at $\sqrt{s_{_{NN}}} = 200$
  GeV~\cite{Adams:2005aw}. Sinusoidal modulations associated with
  $v_2$ have been subtracted. Right panel: $p_T$ autocorrelations
  plotted in cylindrical coordinates. The positive \textit{near-side}
  peak is subtracted revealing a valley~\cite{Trainor:2005ac}.}
\label{autocorr}
\end{figure}

Mishra et. al. propose measurements of the mean square values of flow
coefficients $\sqrt{v_{n}^2}=v_{n}^{rms}$ as a way to test for the
suppression of the long wavelength super-horizon
fluctuations. $v_n^{rms}$ values are related to two particle
correlations which have already been measured by STAR. The
measurements most closely related to the CMB measurements are the
$p_T$ correlations shown in Fig.~\ref{autocorr}. The figure shows
$p_T$ correlations after the $v_2$ like $\cos(2\Delta\phi)$ term has
been subtracted (left) and then after the prominent near-side peak has
been subtracted (right). The subtraction reveals a valley that was
first interpreted as the medium recoiling from an impinging jet. This
interpretation was motivated by the assumption that the near-side peak
is dominated by the fragmentation of hard and semi-hard scattered
partons (mini-jets). We show here that causality can lead to the
observed valley through the suppression of long wavelength,
super-horizon modes.

\begin{figure}[htbp]
\parbox{0.50\textwidth}{
\includegraphics[width=0.5\textwidth]{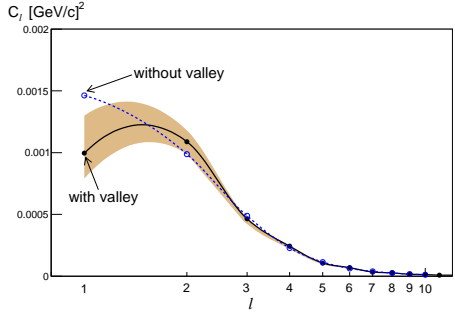}}
\parbox{0.32\textwidth}{
\caption[]{ The power spectrum from \\ 
  $p_T$ fluctuations in heavy-ion collisions. \\
 $C_{l}$ are calculated at midrapidity \\
 with $\theta=0$.}
\label{powerspec} }
\end{figure}

In order to carry through the analogy with the analysis of CMB
temperature fluctuations, we perform a similar power-spectrum analysis
by calculating the coefficients in a spherical harmonic decomposition
of the $p_T$ correlation data. For simplicity we only consider the
transverse direction at $\Delta\eta=0.0$. In ref.~\cite{Adams:2005aw}, the
$p_T$ correlations are fit by a $\cos(\Delta\phi)$, a
$\cos(2\Delta\phi)$, and three two-diminsional Gaussians. In
Fig.~\ref{powerspec} we show the power-spectrum derived from the
Gaussian terms. The spectrum is shown with and without the negative
Gaussian term (i.e. the valley). Note that the existence of the valley
is related to the suppression of the longest wavelength modes (lowest
harmonics). We conclude that the valley is evidence for suppression of
long wavelength fluctuations due to the limits of causality and the
brief duration of heavy-ion collisions.

\section{Conclusions: Life is Short}

The upper limit on the ratio of $\sigma_{v_{2}} / \langle v_2\rangle$
coincides with the eccentricity fluctuations calculated from Monte
Carlo Glauber models of the initial conditions. Removing non-flow
correlations from $\sigma_{v_{2}}$ will reduce the upper limit on the
allowed values of $\sigma_{v_{2}} / \langle v_2\rangle$. Significant
two-particle correlations have been observed. We conclude therefore
that either those correlations are a manifestation of correlations
from the initial conditions, or the Monte Carlo Glauber model
overestimates the fluctuations in the initial conditions.

\begin{wrapfigure}{r}{0.35\textwidth}
\parbox{40mm}{
\vspace{0mm}
  \includegraphics[width=0.33\textwidth]{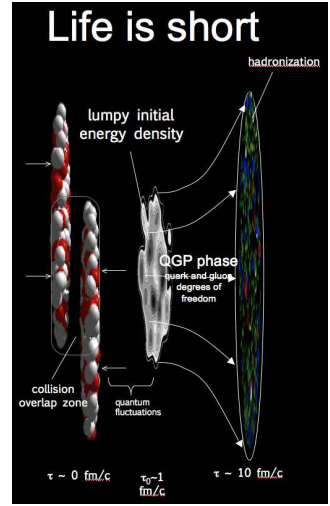}}
\parbox{0.2\textwidth}{
\vspace{0mm}\caption[]{ A proper perspective. \hspace{0.99\textwidth}}}
\label{life}
\end{wrapfigure}

The transfer of spatial correlations and fluctuations from the initial
conditions to correlations in momentum space requires interactions
between constituents. If these interactions do not persist for long
enough, then long wavelength correlations will remain
super-horizon. Given that the diameter of a Au nuclei is approximately
15 fm, if the interactions between constituents produced in a Au+Au
collision only persists for ~10 fm, then regions of the overlap zone
will remain outside the event horizon. This is illustrated in Fig.~\ref{life}.

Causality has been discussed with regards to the longitudinal axis but
it can also play a role in the transverse plane as well. Examining
$p_T$ correlations, we find evidence for suppression of long
wavelength fluctuations due to the limits of causality and the brief
duration of heavy-ion collisions. The suppression of long wave-length,
super-horizon fluctuations can naturally lead to the anti-correlation
observed in $p_T$ correlations which was previously ascribed to recoil
from an impinging hard parton.

\vfill\eject
\end{document}